\documentclass[superscriptaddress,amsfonts,amsmath,amssymb,aps,pre,twocolumn,
floatfix]{revtex4-1}
\usepackage{bm}
\usepackage{graphicx}
\begin{document}
\title{Electron dynamics and injection in plasma-based accelerators with sharp vacuum-plasma transitions}

\author{Ronghao Hu}
\affiliation{State Key Laboratory of Nuclear Physics and Technology, and Key 
Laboratory of HEDP of the Ministry of Education, CAPT, Peking University, 
Beijing 100871, China}
\author{Haiyang Lu}
\email{hylu@pku.edu.cn}
\affiliation{State Key Laboratory of Nuclear Physics and Technology, and Key 
Laboratory of HEDP of the Ministry of Education, CAPT, Peking University, 
Beijing 100871, China}
\affiliation{Collaborative Innovation Center of Extreme Optics, Shanxi University, Taiyuan, Shanxi 030006, China}
\author{Yinren Shou}
\author{Jinqing Yu}
\author{Chia-erh Chen}
\affiliation{State Key Laboratory of Nuclear Physics and Technology, and Key 
Laboratory of HEDP of the Ministry of Education, CAPT, Peking University, 
Beijing 100871, China}
\author{Xueqing Yan}
\email{x.yan@pku.edu.cn}
\affiliation{State Key Laboratory of Nuclear Physics and Technology, and Key 
Laboratory of HEDP of the Ministry of Education, CAPT, Peking University, 
Beijing 100871, China}
\affiliation{Collaborative Innovation Center of Extreme Optics, Shanxi University, Taiyuan, Shanxi 030006, China}

\date{\today}%
\begin{abstract}
The dynamic process of a laser or particle beam propagating from vacuum into underdense plasma has been 
investigated theoretically. Our theoretical model combines a Lagrangian fluid model with the classic 
quasistatic wakefield theory. It is found that background electrons can be injected into wakefields because sharp 
vacuum-plasma transitions can reduce the injection threshold. The injection condition, injection 
threshold as well as the injection length can be given theoretically by our model and are compared with results from 
computer simulations. Moreover, electron beams of high qualities can be produced near the 
injection thresholds and the proposed scheme is promising in reducing the injection threshold and improving the beam 
qualities of plasma based accelerators.
\end{abstract}
\maketitle

Plasma acceleration is a fast-developing technique for accelerating charged particles with plasma waves driven 
by laser or particle beams. Current experiments have shown accelerating gradients several orders of magnitude higher 
than conventional particle accelerators\cite{Esarey2009RevModPhys}, offering a way to build high performance particle 
accelerators of much smaller 
size. The basic concepts of laser wakefield accelerator (LWFA) were first proposed by T. Tajima and J. M. Dawson in 
1979\cite{Tajima1979PhysRevLett}, where electrons are accelerated by plasma waves driven by laser pulses. Several 
years later, P. Chen \textit{et al}\cite{Chen1985PhysRevLett} proposed the concepts of plasma wakefield 
accelerator (PWFA) where high current electron bunches are used to generate intense electric fields to accelerate 
trailing copropagating electron bunches. Experimental 
breakthroughs\cite{Muggli2004PhysRevLett,Hogan2005PhysRevLett,Blumenfeld2007Nature,Litos2014Nature,Esarey2009RevModPhys,
Kim2013PhysRevLett, Wang2013NatComm,Leemans2014PhysRevLett} has been made in recent years 
and electron energy gain was improved to 4.2 GeV in LWFA\cite{Leemans2014PhysRevLett} and over 
40 GeV in PWFA\cite{Blumenfeld2007Nature}. Several injection scheme were 
proposed and investigated, including 
self-injection\cite{Lu2007PhysRevSTAB,Kostyukov2009PhysRevLett,Kalmykov2009PhysRevLett,Froula2009PhysRevLett,
Corde2013NatComm}, downramp 
injection\cite{Geddes2008PhysRevLett,Schmit2010PhysRevSTAB,Buck2013PhysRevLett}, colliding pulse 
injection\cite{Esarey1997PhysRevLett,Faure2006Nature,Davoine2009PhysRevLett,Kotaki2009PhysRevLett,Lehe2013PhysRevLett} 
and 
ionization 
injection\cite{Pak2010PhysRevLett,Clayton2010PhysRevLett,Hidding2012PhysRevLett,Bourgeois2013PhysRevLett,
LiF2013PhysRevLett,Yu2014PhysRevLett, Xu2014PhysRevLett, 
Xu2014PhysRevSTAB,Zeng2015PhysRevLett,Mirzaie2015SciRep}

Despite the great achievements made in the past few decades, we are still far from fully understanding the physics of 
plasma acceleration. For example, the dynamic process of a laser or particle beam propagating from vacuum into 
underdense plasma still lacks investigation. The classic quasistatic wakefield theory treats the drivers as if they 
were born inside the plasmas and the boundary effects are neglected\cite{Esarey2009RevModPhys}. Natural gas plasmas 
produced by gas jets have 
boundary density ramps with lengths from hundreds of microns to millimeters\cite{Hsieh2006PhysRevLett}. 
Quasistatic theory is appliable in 
this case because the density changes slowly as the driver propagates. Several 
techniques\cite{Hsieh2006PhysRevLett,Buck2013PhysRevLett,Helle2016PhysRevLett} have been proposed to produce sharp 
density transitions with transition ramp lengths about tens of microns, which are comparable to the plasma wavelengths 
used for acceleration\cite{Esarey2009RevModPhys} and the wavelengths of mid-infrared lasers (10.6 $\mu$m for CO$_2$ laser). 
For these sharp density transitions, quasistatic theory needs to be modified to comprehend the underlying physics. 

In this letter, we introduce a theoretical model to investigate the dynamic process of a laser or particle beam 
propagating from vacuum into plasma. A Lagrangian fluid model\cite{Dawson1959PhysRev} combined with quasistatic 
wakefield theory can give us a clear physical picture of this dynamic process. We found that the main 
differences between sharp transitions and long transitions are the behaviors of a group of electrons that can enter the 
vacuum and return to the plasma. Some of these refluxing electrons can be injected into the wakefields and 
get accelerated. The dynamics of refluxing electrons and a theoretical derivation of the injection 
condition, injection threshold and injection length of refluxing electron injection (REI) are discussed in the 
following. Theoretical results are compared with numerical results from 1D and 3D particle-in-cell (PIC) simulations 
with EPOCH\cite{Arber2015PlasmaPhysControlFusion}, showing that our model have included the essence of the underlying 
physics. The injection threshold of REI is also found to be smaller than that of 
self-injection\cite{Froula2009PhysRevLett}. The beam qualities of REI are 
characterized with 3D PIC simulations and it 
is found that high quality electron beams can be generated near the injection thresholds.

\begin{figure*}
\centering
\includegraphics[width=1.0\textwidth]{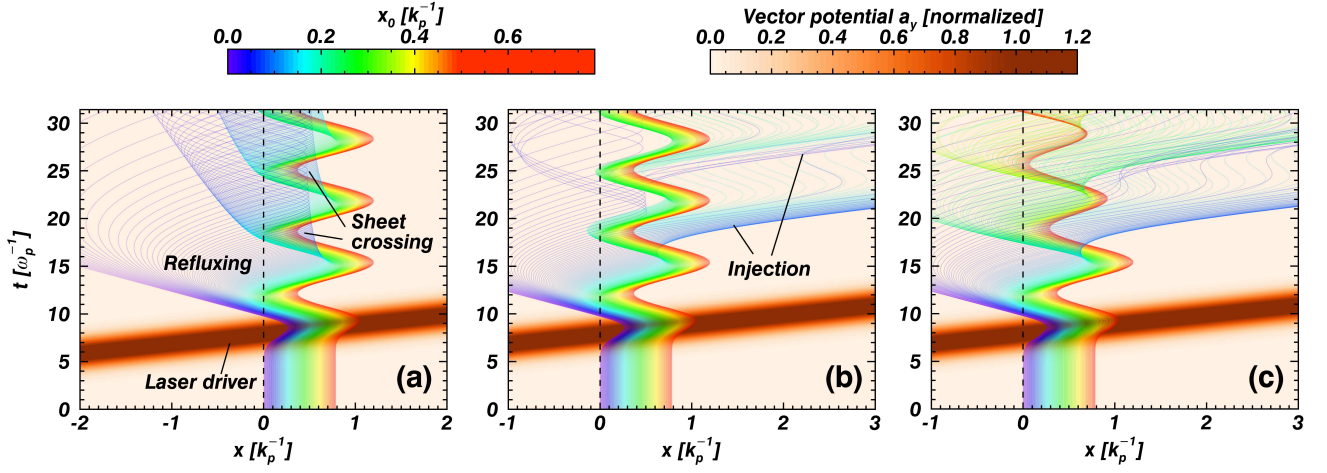}
\caption{Trajectories of electrons with different initial positions marked with different colors. (a) Numerical 
solutions of equations 
(\ref{motion}) without modifications after sheet crossing. (b) Numerical solutions including quasistatic theory 
after sheet crossing. (c) Trajectories obtained from a PIC simulation with the same parameters. The filled contour 
plots in each subfigure show the spatiotemporal profiles of the driving laser and the dashed lines indicate the 
initial plasma boundary. The laser is incident from left to right, and plasma initially locates in the $x>0$ area. 
$a_0$=1.2 and $n_0/n_c$=0.004 are used in all subfigures.}
\label{fig1}
\end{figure*}

For laser drivers and cold plasmas, the 1D equations of motion averaged by laser cycles can be given 
as\cite{Esarey2009RevModPhys}
\begin{equation}
\begin{aligned}
\label{motion}
\frac{dp_x}{dt}=-E+F_p,\ \frac{dx}{dt}=p_x/\gamma.
\end{aligned}
\end{equation}
$t$ is time normalized by $\omega_p^{-1}$, where $\omega_p=\sqrt{n_ee^2/m_e\varepsilon_0}$ is the plasma frequency, 
$n_e$ is plasma density, $e$ is elementary charge, $m_e$ is electron mass at rest and $\varepsilon_0$ is vacuum 
permittivity. $p_x$ is the longitudinal momentum and is normalized by $m_ec$, where $c$ is the light velocity in 
vacuum. 
$x$ is longitudinal position normalized by $k_p^{-1}$, where $k_p=\omega_p/c$ is the plasma wavenumber. $E$ is the 
longitudinal electric field and is normalized by $m_e\omega_pc/e$. $F_p=-\frac{1}{4\gamma}\frac{\partial 
a^2}{\partial x}$ is 
the ponderomotive force of a linearly polarized laser driver, normalized by $m_e\omega_pc$, $a$ is the laser vector 
potential normalized 
by $m_ec/e$ and $\gamma=\sqrt{1+a^2/2+p_x^2}$ is the Lorentz factor. All the physcial quantities are in SI units. For 
particle drivers, one only needs to replace 
the ponderomotive force with the space charge force and neglect the $a^2/2$ term in the Lorentz factor. Before sheet 
crossing happens, the order of electron sheets is not disturbed and the electric force can be obtained from 
Poisson's equation as $E=\int_{-\infty}^x [n_i(x)-n_e(x)]dx=\int_{x_0}^x n_0(x)dx$, where $n_i(x)$ is the positive 
charge density, $n_e(x)$ is electron density and $n_0(x)$ is the initial undisturbed electron 
density\cite{Dawson1959PhysRev}. Here ions are 
assumed to be fixed and all the density values are normalized using $n_0(x=0)$. For simplicity without losing 
generality, the density profile of a vacuum-plasma transition (locates at $x$=0) can be written as $n_0(x)=H(x)$, where 
$H(x)$ is the unit step function, and the electric force for electrons intially at $x_0$ can be written as 
$E=xH(x)-x_0$. 
Equations (\ref{motion}) can be integrated numerically for a linearly polarized gaussian laser pulse with 
$a=a_y=a_0\exp{[-(t-x/v_d-t_{delay})^2/\tau^2]}$, where $a_0$ is the peak vector potential, $v_d$ is the velocity of 
the driver, $t_{delay}$ is the time delay of the pulse peak and $\tau$ is the 1/e half pulse duration. For drivers 
with small amplitudes ($a_0\sim$1), $v_d\approx\sqrt{1-n_0/n_c}$, where $n_c$ is the critical plasma 
density\cite{Esarey2009RevModPhys}. Fig. \ref{fig1}(a) shows the numerical solutions of equations (\ref{motion}), 
where the rainbow-colored wavy lines are particle trajectories moving in the laser and charge separation fields, with 
colors being the initial positions of the particles, $x_0$, given by the color bar on top. Some particles (like red 
ones in Fig. \ref{fig1}(a)) oscillate and remain approximately the same region in the plasma while the laser moves 
forward into the plasma. Electrons near the boundary (the blue ones in the back) are ejected into the vacuum, and sheet 
crossing\cite{Dawson1959PhysRev}, or the crossing of electron trajectories happens shortly after they re-enter the 
plasma. 
Electrons initially deep inside the plasma are oscillating with constant amplitudes and by neglecting the ponderomotive 
force term in equations (\ref{motion}), one can obtain a constant of motion for the oscillation, i.e. 
$\mathcal{H}=\sqrt{1+p_x^2}-\Psi$, where 
$\Psi$ is the electrostatic potential and $\partial \Psi/\partial x=-E$. An approximate solution for small amplitude 
oscillation can be obtained as
\begin{equation}
\begin{aligned}
\label{solution}
&p_x(t,x_0)=-p_m\cos{[\omega_\gamma(t-x_0/v_d-t_0)]},\\
&x(t,x_0)\ 
=x_0-\delta_m\sin[\omega_\gamma(t-x_0/v_d-t_0)].
\end{aligned}
\end{equation}
$p_m$ is the maximum oscillation momentum, $\gamma_m=\sqrt{1+p_m^2}$ is the maximum Lorentz factor after the driver and 
$\delta_m=\sqrt{2(\gamma_m-1)}$ is the maximum displacement of the oscillation. $t_0$ is the time electron with $x_0=0$ 
enters the vacuum. For gaussian drivers with $a_0\sim1$, $\tau\approx\pi/2$ and $v_d\sim1$, we found 
$\gamma_m\approx1.005+0.166(a_0-0.5)^2$. $\omega_\gamma$ is the relativistic oscillation frequency, and can be fitted 
as $\omega_\gamma\approx\gamma_m^{-0.44}$.
For electrons have initial positions $0<x_0<\delta_m$, they follow the same oscillation motion until they reach the 
plasma boundary and then enter the vacuum ($x<0$), where they experience constant positive 
electric forces as $E(x)=-x_0$. By this force, they will return to the plasma with a maximum momentum equal to $p_m$, 
as shown in Fig. \ref{fig1}(a). 
The refluxing time, which is defined as the time from the 
refluxing electron leaving its initial position with negative velocity to its returning, can be written as 
$t_{re}=2\arccos(p_0/p_m)/\omega_\gamma+2p_0/x_0$, where $p_0=\sqrt{(\gamma_m-x_0^2/2)^2-1}$ is the electron momentum 
at $x=0$. As one can see, the refluxing time $t_{re}$ is a function of the initial position $x_0$ and its value ranging 
from $\pi/\omega_\gamma$ to $+\infty$. For refluxing electrons with proper refluxing time, their trajectories will 
intersect with the trajectories of oscillating electrons (Fig. \ref{fig1}(a)). The trajectories of refluxing electrons 
inside the plasma and the trajectory of the outermost oscillating electrons ($x_0=\delta_m$) will cross on condition 
that
\begin{equation}
\begin{aligned}
\label{crossing}
&x_0+\delta_m\sin[\omega_\gamma(t_{sc}-x_0/v_d-t_{re}-t_0)]\\
&=\delta_m-\delta_m\sin[\omega_\gamma(t_{sc}
-\delta_m/v_d-t_0) ] .
\end{aligned}
\end{equation}
Solving equation (\ref{crossing}), one can obtain the sheet crossing 
time $t_{sc}=t_0+x_0/v_d+t_{re}+\arcsin[A/\sqrt{2+2\sin(B)}-C]/\omega_\gamma$, where $A=1-x_0/\delta_m$, 
$B=\omega_\gamma[(\delta_m-x_0)/v_d-t_{re}]$, $C=\arcsin[\cos(B)/\sqrt{2+2\sin(B)}]$. 

Sheet crossing will always happen 
near the vacuum-plasma transition, but only when the plasma waves are strong enough can refluxing electrons be trapped. 
After sheet crossing, due to the disordering of the electrons, it will be difficult to obtain the electric fields 
analytically. To find out the injection condition and injection threshold, we neglect the beam loading effects of the 
refluxing electrons after they cross with the outermost oscillating electron, i.e. they do not contribute to the 
electric field and disturb the motions of oscillating electrons. We can make this assumption because near the 
injection threshold, the injected charge is small and the beam loading effects are not significant. With this 
assumption, the electric force of 
the refluxing electrons can be obtained using the quasistatic theory\cite{Esarey2009RevModPhys} and their trajectories 
can be computed after sheet crossing as shown 
in Fig. \ref{fig1}(b). As a reference, the trajectories obtained from a 1D PIC simulations with the same parameters are 
shown in Fig. \ref{fig1}(c), and one can see the dynamics of the injected electrons are much alike despite the 
approximations we have made. Refluxing electrons can be injected into plasma waves if they are above the separatrix of 
injection when they cross with the outermost oscillating electron\cite{Esarey2009RevModPhys,Faure2016CERN}. The 
momentum 
of the outermost oscillating electron at sheet crossing moment is 
$p_{os}=-p_m\cos[\omega_\gamma(t_{sc}-\delta_m/v_d-t_0)]$, the separatrix of injection can be written as 
$p_{sp}=\left[v_dD-\sqrt{D^2-(1-v_d^2)}\right]/(1-v_d^2)$, where 
$D=\sqrt{1-v_d^2}-\sqrt{1+p_m^2}+\sqrt{1+p_{os}^2}+v_d(p_m-p_{os})$. The momentum of refluxing electron at sheet 
crossing time $t_{re}$ is $p_{re}=p_m\cos[\omega_\gamma(t_{sc}-x_0/v_d-t_{re}-t_0)]$, and the injection condition can 
be 
thus written as 
\begin{equation}
\label{condition}
p_{re}-p_{sp}>0.
\end{equation} 

\begin{figure*}
\centering
\includegraphics[width=0.75\textwidth]{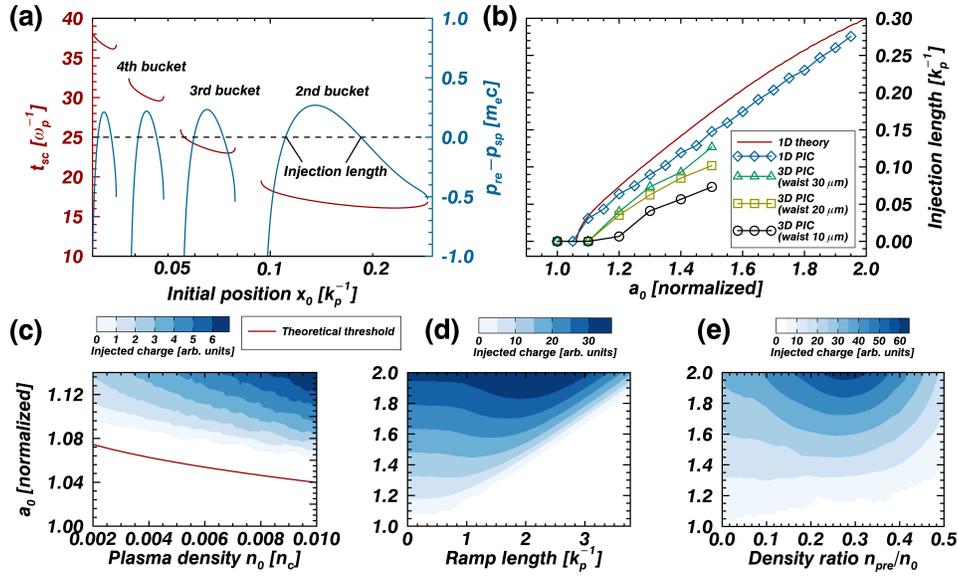}
\caption{(a) Refluxing time $t_{sc}$ and momentum difference $p_{re}-p_{sp}$ as a function of electron initial 
position $x_0$. The laser $a_0$ is 1.2. (b) Injection length of second bucket for different laser peak amplitude $a_0$. 
(c) Injected charge in 
second bucket for different $a_0$ and plasma density $n_0$. The red line is the theoretical threshold calculated by solving $max(p_{re}-p_{sp})=0$. (d) Injected charge in second bucket for different $a_0$ 
and boundary ramp length. (e) Injected charge in second bucket for different $a_0$ and density ratio of a transition 
from plasma with density $n_{pre}$ to plasma with density $n_0$.  Plasma density $n_0/n_c$=0.004 is used for (a), (b), (d) and (e).}
\label{fig2}
\end{figure*}

The sheet crossing time $t_{sc}$ and momentum difference $p_{re}-p_{sp}$ against 
different 
initial position $x_0$ are plotted in Fig. \ref{fig2}(a). As one can see, there are several separated injection areas 
with 
$p_{re}-p_{sp}$ above zero, and according to the corresponding $t_{sc}$ (with the time reference of Fig. \ref{fig1}), 
we 
can tell that electrons from these injection area are loaded into different acceleration buckets of the 
wakefields. Different from self-injection and downramp injection\cite{Esarey2009RevModPhys,Faure2016CERN}, REI can not 
load electrons into the first acceleration bucket. The second 
acceleration bucket has the longest injection length and highest injected charge. Theoretical 
injection length of the second bucket as a function of normalized laser amplitude $a_0$ is shown in Fig. \ref{fig2}(b) 
together with numerical results from 1D and 3D PIC simulations with different laser waists (gaussian transverse 
profiles). The 
injection threshold, or the minimum $a_0$ required for injection, is about 1.05 according to the theory and 1D PIC 
simulations. The 1D PIC injection length is smaller than the theory for $a_0$ larger than the injection threshold. This 
is caused by the beam loading effects of injected bunch, which we neglected in our theoretical model. Due to the beam 
loading, the amplitudes of wakefields will be reduced and injected charge will be limited. The injection thresholds of 
3D PIC simulations are larger than that of the 1D theory and simulation and with the decrease of the laser waist, the 
threshold increases. Because of the transverse ponderomotive force of the guassian pulse, electrons are pushed away 
from the laser axis and the electric fields experienced by electrons are actually smaller than the on-axis value, which 
increases the threshold for injection. 
According to the 3D simulations, the power threshold for REI is $P/P_c\sim1.4$ (varies for different densities, 
$P_c\simeq17.5n_c/n_0$ [GW] is the critical power for self-focusing), which is much smaller than that of self-injection 
measured at similar plasma densities ($P/P_c\sim3$)\cite{Froula2009PhysRevLett}. For typical plasma densities used in 
LWFA ($10^{18}\sim10^{19}$ cm$^{-3}$), 
the injection threshold increases with the decrease of the plasma density, but the variations are small as shown in 
Fig. \ref{fig2}(c). For more realistic vacuum-plasma transitions, we can model the density profile with 
$n_0(x)=H(x+l)H(-x)(x+l)/l+H(x)$, where $l$ is the length of a linear density ramp. For small ramp lengths, the 
injection thresholds are similar to the ideal case we discussed above, as depicted in Fig. \ref{fig2}(d). When ramp 
length is longer than $k_p^{-1}$, the injection threshold almost linearly increases with the ramp length (Fig. 
\ref{fig2}(d)). For the ideal case, the refluxing electrons are injected into quasistatic 
plasma waves (wakes), the phase velocity of the wakes is approximately the group velocity of the laser. But for the 
long ramp cases, the refluxing electrons are injected into plasma waves with decreasing wavelengths. The equivalent 
phase velocity of the wakes can be much larger than light velocity and the injection thresholds are larger than the 
ideal case. When the transition is not from vacuum to plasma but from a low density plasma to a higher density one, the injection 
threshold also increases with the density ratio of the transition, as shown in Fig. \ref{fig2}(e). The injection threshold of the third 
bucket also changes due to the change of the ramp length or density ratio. Typically, the injection threshold of the third bucket is 
smaller than that of the second if the ramp length or density ratio is larger than zero.

\begin{figure}
\centering
\includegraphics[width=0.45\textwidth]{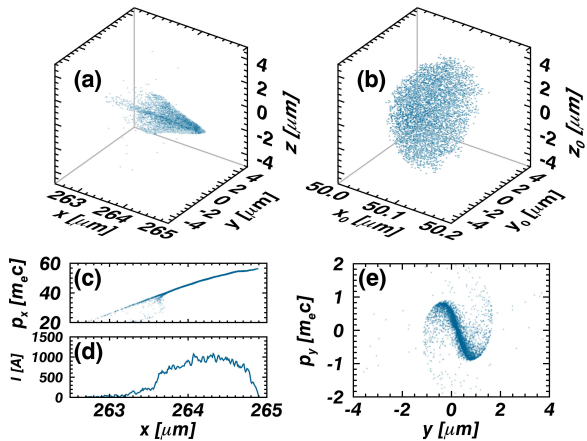}
\caption{3D PIC simulation results of the REI electron bunch after propagating about 200 $\mu$m in plasma. The 
electrons are the same group of electrons in Table \ref{tab1} with $a_0$=1.4, $w$=10 $\mu$m. (a) Positions of injected 
electrons. (b) Initial positions of injected electrons. The plasma boundary 
locates at $x=50$ $\mu$m. (c) Longitudinal phasespace distribution of injected electrons. (d) Current profile of the 
injected bunch. (e) Transverse (along the laser polarization direction) phasespace distribution of the injected 
electrons. The wavelength of driving laser is 800 nm and the plasma density is 0.004$n_c$.}
\label{fig3}
\end{figure}

The beam qualities are determined by the electron distributions in six-dimensional phasespace 
($x,y,z,p_x,p_y,p_z$)\cite{Di2014PhysRep}. 
The initial distributions of injected electrons in real space, or the collection volume must be well-confined to 
produce beams of high qualities\cite{Corde2013NatComm}. For REI, the electron distribution after injection and the 
collection 
volume are shown in Figs. \ref{fig3}(a) and (b), respetively. The longitudinal bunch length is one order of 
magnitude larger than the injection length because of the differences in refluxing time for different inital positions. 
The collection volume will 
increase (in larger radius and longer injection length) with the increase of the laser amplitude or waist. The 
injected charge and current will also increase (Table \ref{tab1}). To control the collection volume, it is favorable to 
operate the injection near the threshold. The injected bunch is negative chirped but the 
slice energy spread is very low, as shown in Fig. \ref{fig3}(c) and Table \ref{tab1}. The front of the bunch are 
inside the focusing part of the wakefields and can preserve the beam qualities in 
acceleration\cite{Esarey2009RevModPhys}. Electrons in the rear 
are inside the defocusing part and they will eventually leave the acceleration region. The current profile is shown in 
Fig. \ref{fig3}(d), which is close to a flattop profile. Before 
sheet crossing, the transverse momentums of refluxing electrons are small ($\sim$0.1 $m_ec$)\cite{SM}. And when 
electrons are 
injected into the wakes, they start 
the well-known betatron oscillations due to the transverse focusing forces of the wakes\cite{Xu2014PhysRevLett}. 
Because the injection happens in a period of time ($\sim\omega_p^{-1}$) much smaller than the period of betatron 
oscillations ($\sim\sqrt{\gamma}\omega_p^{-1}$)\cite{Xu2014PhysRevLett}, the betatron phases of the injected electrons 
will be similar (Fig. \ref{fig3}(e)), which results in pretty small transverse emittances (0.1-1 $\mu$m) and high 
brilliances ($\sim$10$^{16}$ A/m$^2$), as shown in Table 
\ref{tab1}. The overall beam qualities of REI bunches are comparable to the beams driving free 
electron lasers\cite{Di2014PhysRep}.
\begin{table}
\resizebox{0.45\textwidth}{!}{
\begin{tabular}{|c|r|r|r|r|r|r|r|r|r|}
\hline $w$ [$\mu$m] &      \multicolumn{3}{c|}{10} &      \multicolumn{3}{c|}{20} &      \multicolumn{3}{c|}{30} \\
\hline $a_0$        &     1.3 &     1.4 &      1.5 &    1.3 &       1.4 &     1.5 &      1.3 &     1.4 &     1.5 \\
\hline $B_{n}$ 
[10$^{16}$ A/m$^2$] &   0.474 &    1.50 &     3.68 &   16.7 &      3.97 &    1.27 &     2.55 &    1.51 &   0.780 \\
\hline $I$ [kA]     &   0.491 &    1.12 &     1.92 &   3.38 &      6.05 &    7.31 &     8.45 &    14.9 &    17.6 \\
\hline $Q$ [pC]     &    1.19 &    3.67 &     7.36 &   11.0 &      24.5 &    40.3 &     27.2 &    55.9 &    90.7 \\
\hline $\tau$ [fs]  &    2.60 &    3.87 &     4.74 &   4.07 &      5.07 &    6.80 &     3.87 &    4.74 &    5.94 \\
\hline $\varepsilon_
{ny}$ [$\mu$m]      &   0.466 &   0.361 &    0.338 &  0.231 &     0.641 &    1.17 &    0.934 &    1.51 &    2.13 \\
\hline $\varepsilon_
{nz}$ [$\mu$m]      &   0.444 &   0.413 &    0.308 &  0.175 &     0.477 &   0.980 &    0.710 &    1.31 &    2.11 \\
\hline $\sigma_
{\gamma s}$ [MeV]   &   0.200 &   0.118 &    0.214 &  0.275 &     0.660 &    1.14 &    0.631 &    1.15 &    1.64 \\
\hline
\end{tabular}}
\caption{Beam parameters of REI after propagating about 200 $\mu$m in plasma obtained from 3D PIC simulations. $w$ is 
the laser waist, $B_{n}$ is the peak 
brilliance, $I$ is the peak current, $Q$ is the injected charge (with $p_x/m_ec$ above 20) in the second bucket, $\tau$ 
is the FWHM bunch duration, $\varepsilon_{ny}$ and $\varepsilon_{nz}$ are normalized emittances in 
$y$ and $z$ directions, $\sigma_{\gamma s}$ is 
the averaged RMS slice energy spread. The slice length is chosen as 20 nm, and the slice energy spread is averaged over 
all slices 
within the FWHM of the current profile. The wavelength of driving laser is 800 nm and the plasma density is 
0.004$n_c$.}
\label{tab1}
\end{table}

REI is easier to be realized in lower density plasmas, because the required boundary ramp length can be larger. 
The beam qualities of REI can be well-preserved in the following acceleration with the scheme using different 
plasma densities for injection and acceleration\cite{Wang2016PhysRevLett}. Manipulating the energy chirp of the 
injected bunch\cite{Brinkmann2017PhysRevLett,Hu2016PhysRevAB}, we can also control the energy spread of the beam. REI can also 
be applied in a staged laser plasma accelerator\cite{Steinke2016Nature}, considering the small threshold required for 
injection. 

In summary, the dynamic process of a laser or particle beam propagating from vacuum into underdense plasma has 
been investigated theoretically. With a Lagrangian fluid model combined with the 
classic quasistatic wakefield theory, we found that background electrons can be injected into wakefields because 
sharp vacuum-plasma transitions can reduce the injection threshold. The injection condition, injection 
threshold as well as the injection length can be given theoretically by our model and are compared with results from 
computer simulations. Electron beams of high qualities can be produced near the 
injection thresholds and the proposed scheme is promising in reducing the injection threshold and improving the beam 
qualities of plasma based accelerators. 

\begin{acknowledgments}
The authors want to thank Antonio Ting for useful discussion. The PIC simulatons were carried out in Shanghai Super 
Computation Center and Max Planck Computing and Data Facility. 
This work was supported by National Basic Research Program of China (Grant No. 2013CBA01502), National Natural Science 
Foundation of China (Grant Nos. 11575011,11535001) and National Grand Instrument Project (2012YQ030142). The EPOCH 
program was funded by the UK EPSRC grants EP/G054950/1, EP/G056803/1, EP/G055165/1 and EP/M022463/1. J. Yu wants to 
thank the Project 2016M600007 funded by China Postdoctoral Science Foundation.
\end{acknowledgments}

\end{document}